\documentclass{elsart}
\usepackage{epsfig}
\begin{document}
\runauthor{Vassiliev et al.}
\begin{frontmatter}
\title{Flux Sensitivity of VERITAS}
\author{V.\ V.\ Vassiliev}

\address{Whipple Observatory, Harvard-Smithsonian CfA, 
P.O. Box 97, Amado, AZ 85645, USA \\ For the VERITAS Collaboration}

\begin{abstract}
VERITAS is a new major ground-based gamma-ray observatory with an array of 
seven $10$m optical reflectors to be built at the Whipple Observatory in 
southern Arizona, USA. It will consist of an array of imaging Cherenkov 
telescopes designed to conduct critical measurements of AGNs and SNRs in the 
energy range of $50$ GeV - $50$ TeV. The design of the array has been 
optimized for the highest sensitivity to point sources in the 
$100$ GeV - $10$ TeV band when the stereoscopic imaging technique is 
employed. Maximum versatility of the array has been another major 
optimization criterion. We present the flux sensitivity of the baseline 
VERITAS configuration.
\end{abstract}



\end{frontmatter}

\section{Introduction}

A new generation of very high energy ground based
$\gamma$-ray observatories (VERITAS \cite{VERITASLOI}, 
HESS \cite{HESS}, MAGIC \cite{MAGIC}) 
promises to extend their sensitive energy range to 
below $100$ GeV, an energy band which is expected to   
contain a wealth of new information on high energy physics
and astrophysics (for review see \cite{Weekes}). 
The spectra and variability of AGNs, the origin of
high energy cosmic rays, physical processes in strong magnetic 
fields of pulsars, the puzzle of dark matter, the type of 
cosmology of the Universe, origin of $\gamma$ ray bursts,
evaporation of black holes, and even a quantum gravity observable
effects, are all in the scope of phenomena which will be studied
by these projects. In this paper we give the technical characteristics 
of VERITAS driven by scientific goals. Also, the main 
motivations for choices of the array parameters are explained. 
Finally, we discuss factors which limit the flux sensitivity.

\section{Simulations}

To predict the performance of VERITAS, the response of the array 
has been simulated with the use of the DePauw-Purdue KASCADE system of 
air shower Monte Carlo (MC) programs \cite{Kascade}. The study of 
hadronic showers which may mimic pure electromagnetic cascades has been
done with the CORSIKA code \cite{Corsika}. 
To determine the optimum VERITAS design we have studied
the effects of varying the number of telescopes 
in an array, the spacing between them, reflector aperture, telescope 
focal length, camera Field of View (FoV) and pixel size. The 
characteristics of the baseline configuration  as well as simulation
input parameters are summarized in Table~\ref{tab1}.

\begin{table}[hb]
\caption{\label{tab1} Specifications of the baseline VERITAS design.}
\begin{center}
\renewcommand{\arraystretch}{1.1}
\begin{tabular}{r|l} \hline\hline
{\bf Location}& Montosa Canyon, Arizona, USA \\
{\bf Array elevation}& $1390$ m  a.s.l. \\
{\bf Number of telescopes}& $7$ (hexagonal layout)\\
{\bf Telescope spacing}& $80$ m  \\
{\bf Mirror}& Davies-Cotton \\
{\bf Reflector aperture/area}& $10$ m / $78.6$ m$^2$\\
{\bf Focal length}& $12$ m \\
{\bf Facets }& $244$, $61$ cm hexagon \\
{\bf Camera }& Homogeneous \\
{\bf Field of View }& $3.5$ deg \\
{\bf Number of pixels }& $499$   \\
{\bf Pixel Spacing/Photocathode Size }& $0.148$ deg/ $0.119$ deg \\
{\bf ADC integration gate }& $8$ nsec  \\
{\bf Pixel BS noise }& $1.1$  pe pixel$^{-1}$ \\
{\bf Weathering reflectivity factor }& $0.9$ \\
{\bf Light-concentrator enhancement}& $1.35$ \\
{\bf Telescope Triggers }& $2,3$ pixels (adjacent) \\
{\bf Pixel coincidence gate }& $15$ nsec \\
{\bf Array Trigger }& $3$ telescopes out of $7$ \\
{\bf Telescope coincidence gate }& $40$ nsec \\ \hline
\end{tabular}
\end{center}
\end{table}

\section{VERITAS design}

The VERITAS design has been optimized for maximum sensitivity 
to point sources in the energy range $100$ GeV - $10$ TeV,
but with significant sensitivity in the range $50$ GeV - $100$ GeV
and from $10$ TeV to $50$ TeV. Optimization has been performed 
with fixed total number of channels.
The details of the VERITAS design study are given in \cite{VERITAS}.
Here we summarize our arguments for the baseline configuration. 
\begin{itemize}
\itemsep=0pt
\item VERITAS should have at least $6-7$ telescopes for good wide 
energy range sensitivity and versatility. Seven telescopes provide more 
flexibility in VERITAS operation modes when the array is split 
into sub-arrays. The array of three 15m telescopes is rejected because 
of poor performance at high energies and lack of versatility.
\item Camera FoV, $3.5^{\circ}$, is a compromise between achieving 
low energy threshold of the array and array performance at high energies, 
and the ability to conduct an efficient sky survey and study 
extended sources. Also, the chosen FoV is the minimum necessary
for effective image reconstruction with a single telescope.
\item The given FoV together with the number of channels per telescope, $500$, 
translates into a pixel spacing of $0.149^{\circ}$.
\item The optical system of the telescope is proposed to be $f/1.2$.
This provides an adequate match to the number of channels in the telescope
camera making reflector global aberrations at the edge of the FoV
comparable to the pixel size. The $f/1.5$ system would perform better
but it would require a substantial additional investment in an 
optical support structure and system of mirror alignment.  
\item The telescope aperture, $10$ m, is chosen by previous successful
experience of operating the Whipple Observatory telescope and for economy.
\item Spacing between telescopes should be in the range $70-80$ m.
Decrease of the spacing reduces efficient event reconstruction, 
background rejection and array sensitivity in the range $200$ GeV - $1$ TeV. 
Increasing the spacing, does not change array sensitivity in this 
energy range, but it increases the array energy threshold. 
\end{itemize}

\section{VERITAS flux sensitivity}

The VERITAS flux sensitivity has been estimated for $50$ hours
of observations on a point source with a spectrum given by 
$dN_{\gamma}/dE \propto E^{-2.5}$ as is seen from the Crab Nebula in the 
sub-TeV energy range \cite{Hillas}. The minimum detectable flux of 
$\gamma$-rays is constrained by the $5\sigma$ confidence level or by the
statistics of the detected photons, $N_{\gamma} > 10$, when the 
background is almost negligible. The details of the sensitivity 
calculations can be found in \cite{VERITAS}.

\begin{figure}
\epsfig{file=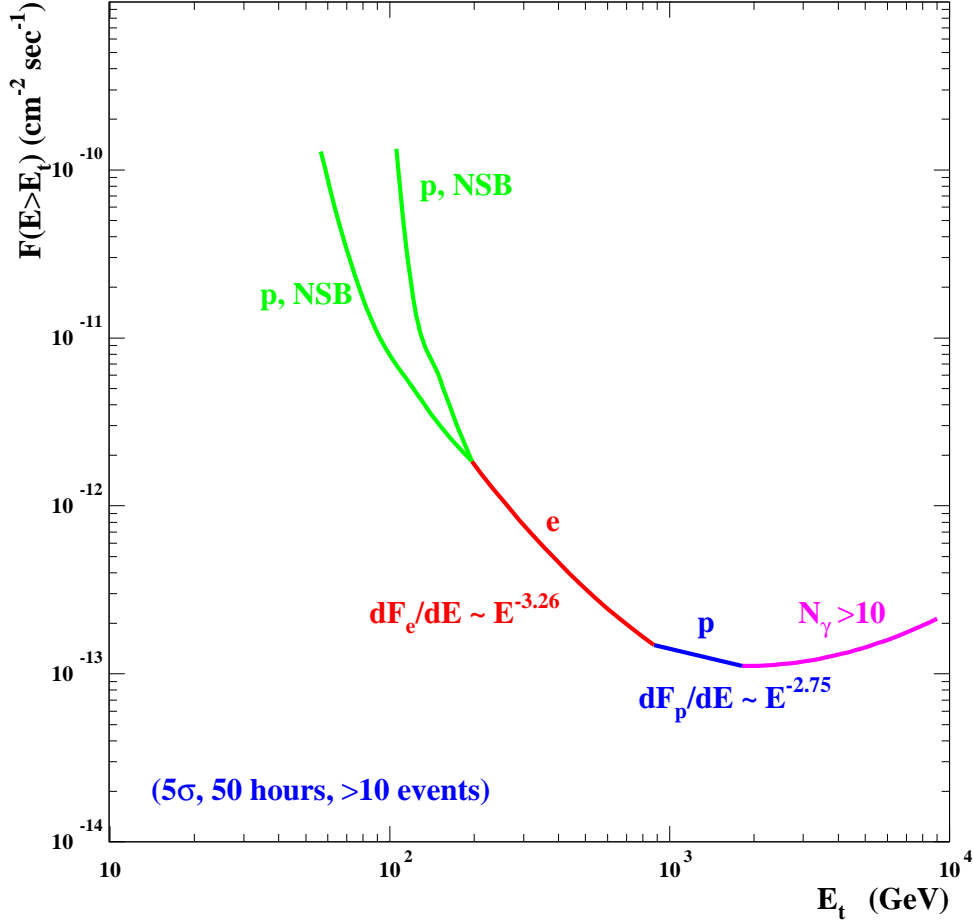,width=5.5in}
\caption{\it The sensitivity of VERITAS to a point-like source in 50 hours 
of observing. The dominant background for VERITAS as a function of 
energy threshold is indicated (see text for details).
\label{fig1}}
\end{figure}
The VERITAS $\gamma$-ray flux sensitivity as a function of array
energy threshold is shown in Figure~\ref{fig1}. At high energies, 
above $\sim 2-3$ TeV, VERITAS sensitivity is limited by photon 
statistics. In this region sensitivity is depressed with energy 
increase due to a limited FoV of the telescope camera. Large zenith 
angle observations can improve array performance in this energy band.
In the vicinity of $1$ TeV, VERITAS sensitivity will likely be 
limited by cosmic ray (CR) protons which mimic $\gamma$-ray showers. 
The detection rate of this isotropic background is not well known
and it may be of scientific interest by itself because of the 
large uncertainty in its predicted effect by different proton interaction 
models available for study within the CORSIKA code. The energy region 
from $200$ GeV to $\sim 1$ TeV will most likely be dominated by 
the diffuse electron background. The steepness of the CR electron 
spectrum causes this to become a limiting factor of the VERITAS 
sensitivity. The region below $\sim 200$ GeV is strongly affected by the night
sky background (NSB) and CR protons. Two curves in this region show
the difference in array sensitivity depending on the conditions 
of the observations. In a favorable situation, dark observation field, 
the array energy threshold for events which we are able to reconstruct 
may decrease to as low as $40-50$ GeV. The single telescope accidental trigger 
rate ($< 0.1-1.0$ MHz), however, may limit  array operation energy threshold 
to $\sim 70$ GeV. If observations are carried out in a bright region of sky 
(Milky Way, lower elevation) where the NSB is $\sim 4$ times brighter, 
VERITAS may be limited to a $110$ GeV energy threshold. The result of 
observations in this energy band will be highly sensitive to array trigger 
condition, and event reconstruction and background rejection methods. 
The plot is indicative of our current achievement, which will certainly 
be improved.

\end{document}